\input{aipcheck}

\documentclass[
    ,final            
  ]
  {aipproc}

\layoutstyle{8x11single}

\begin{document}

\title{Astrophysical Implications of the Superstring-Inspired
$E_6$ Unification and Shadow Theta-Particles}


\author{C.R. Das}{
  address={Centre for Theoretical Particle Physics, Technical
University of Lisbon,\\ Avenida Rovisco Pais, 1, 1049-001 Lisbon,
Portugal}
}

\author{L.V. Laperashvili}{
  address={The Institute of Theoretical and Experimental Physics,\\
Bolshaya Cheremushkinskaya, 25, 117218 Moscow, Russia}
}

\author{A. Tureanu}{
  address={Department of Physics, University of Helsinki and Helsinki
Institute of Physics,\\ P.O.Box 64, FIN-00014 Helsinki, Finland}
}

\begin{abstract}
We have developed a concept of parallel existence of the ordinary
(O) and mirror (M), or shadow (Sh) worlds. $E_6$ unification,
inspired by superstring theory, restores the broken mirror parity
at the scale $\sim 10^{18}$ GeV. With the aim to explain the tiny
cosmological constant, we consider the breakings: $E_6 \to
SO(10)\times U(1)_Z$ -- in the O-world, and $E'_6 \to SU(6)'\times
SU(2)'_{\theta}$ -- in the Sh-world. We assume the existence of
shadow $\theta$-particles and the low energy symmetry group
$SU(3)'_C\times SU(2)'_L\times SU(2)'_{\theta}\times U(1)'_Y$ in
the shadow world, instead of the Standard Model. The additional
non-Abelian $SU(2)'_{\theta}$ group with massless gauge fields,
``thetons'', has a macroscopic confinement radius
$1/\Lambda'_{\theta}$. The assumption that
$\Lambda'_{\theta}\approx 2.3\cdot 10^{-3}$ eV explains the tiny
cosmological constant given by recent astrophysical measurements.
Searching for the Dark Matter (DM), it is possible to observe and
study various signals of theta-particles.
\end{abstract}

\maketitle


\section{Introduction}

The present talk is devoted to the problem of cosmological
constant. Our model is based on the following assumptions:
\begin{itemize}
\item Grand Unified Theory (GUT) is inspired by Superstring
theory \cite{1}, which predicts $E_6$ unification in the
4-dimensional space, occurring at the high energy scale
$\sim 10^{18}$ GeV.
\item There exists a Mirror World (MW) \cite{2,3}, which is a
mirror duplication of our Ordinary World (OW), or Shadow World
(ShW) (hidden sector) \cite{4,5}, which is not identical with the
O-world, having different symmetry groups. The mirror (M), or
shadow (Sh) matter interacts with ordinary matter only via
gravity, or other very weak interactions.
\item The Shadow world is responsible for the dark energy (DE)
and dark matter (DM).
\item We assume that $E_6$ unification had a place in the O-
and M- worlds at the early stage of our Universe. This means that
at very high energy scale $\sim 10^{18}$ GeV the mirror
world exists and the group of symmetry of the universe is
$E_6\times E'_6$ \cite{6,7} (where the superscript `prime' denotes
the M- or Sh-world).
\end{itemize}

\section{Mirror world with broken mirror parity}

At low energies we can describe the ordinary and mirror worlds by 
a minimal symmetry $G_{SM}\times G'_{SM},$ where $G_{SM} =
SU(3)_C\times SU(2)_L\times U(1)_Y$ stands for the Standard Model
($SM$) of observable particles: three generations of quarks and
leptons and the Higgs boson. Then $G'_{SM} = SU(3)'_C\times
SU(2)'_L\times U(1)'_Y$ is its mirror gauge counterpart having
three generations of mirror quarks and leptons and the mirror
Higgs boson. The M-particles are singlets of $G_{SM}$ and the
O-particles are singlets of $G'_{SM}$. If the ordinary and mirror
worlds are identical, then O- and M-particles should have the same
cosmological densities. But this is in the immediate conflict with
recent astrophysical measurements. Mirror parity (MP) is not
conserved \cite{8}. In the case of the broken MP the VEVs of the
Higgs doublets $\Phi$ and $\Phi'$: $\langle\Phi\rangle=v$,
$\langle\Phi'\rangle=v'$ are not equal: $v\neq v'$. We have
introduced the parameter characterizing the violation of MP: $
\zeta = v'/v \gg 1$. Then the masses of fermions and massive
bosons in the mirror world are scaled up by the factor $\zeta$
with respect to the masses of their counterparts in the ordinary
world: $m'_{q',l'} = \zeta m_{q,l}$, $M'_{W',Z',\Phi'} =
\zeta M_{W,Z,\Phi}$, while photons and gluons remain massless in
both worlds.

In the language of neutrino physics, the O-neutrinos
$\nu_e$, $\nu_{\mu}$, $\nu_{\tau}$ are \underline{active} neutrinos,
while the M-neutrinos $\nu'_e$, $\nu'_{\mu}$, $\nu'_{\tau}$ are
\underline{sterile} neutrinos. If MP is conserved ($\zeta = 1$), 
then the neutrinos of the two sectors are strongly mixed. But it 
seems that the situation with the present experimental and 
cosmological limits on the active-sterile neutrino mixing do not 
confirm this result. MP is spontaneously broken, $\zeta \gg 1$, 
and active-sterile mixing angles should be small:
$\theta_{\nu\nu'}\sim \frac 1{\zeta}$. Then we have the following
relation between the masses of the light left-handed neutrinos:
$m'_{\nu}\approx \zeta^2 m_{\nu}$. Also the seesaw mechanism
described by Refs.~\cite{8} predicts that so called right-handed
neutrinos $N_a$ with large Majorana mass terms have equal masses
in the O- and M(Sh)-worlds: $M'_{\nu,a} = M_{\nu,a}$. They are
created at seesaw scale $M_R$ (or $M'_R$) in the O- (or
M(Sh)-)world. And even in the model with broken MP, we have the
same seesaw scales in both worlds: $M'_R = M_R$.

\section{Superstring theory and $E_6$ unification}

The `heterotic' superstring theory $E_8\times E'_8$ was suggested 
as a more realistic model for unification of all gauge interactions
with gravity \cite{1}. This ten-dimensional Yang-Mills theory can
undergo spontaneous compactification. The integration over six
compactified dimensions of the $E_8$ superstring theory leads to
the effective theory with the $E_6$ unification in the
four-dimensional space.

In the present investigation at the scale $\sim 10^{18}$ GeV we 
adopt for the O-world the breaking  $E_6\to SO(10)\times U(1),$
while for the Sh-world we consider the breaking $E'_6\to
SU(6)'\times SU(2)',$ thus being able to explain the small value
of the cosmological constant $CC$, due to the additional $SU(2)'$
gauge symmetry group appearing in the Sh-world, which has a large
confinement radius.

We assume that in the ordinary world, from the $SM$ up to the
$E_6$ unification, there exists the following chain of symmetry
groups:
$$SU(3)_C\times SU(2)_L\times U(1)_Y\to  
\left[SU(3)_C\times SU(2)_L\times U(1)_Y\right]_{SUSY}$$
$$\to SU(3)_C\times SU(2)_L \times SU(2)_R\times U(1)_X\times U(1)_Z$$ 
\begin{equation} \to SU(4)_C\times SU(2)_L \times SU(2)_R\times U(1)_Z
\to SO(10)\times U(1)_Z\to E_6. \label{A} \end{equation} 

We consider the
following chain of possible symmetries in the Sh-world:
$$SU(3)'_C\times SU(2)'_L\times SU(2)'_{\theta}\times U(1)'_Y \to 
\left[SU(3)'_C\times SU(2)'_L\times SU(2)'_{\theta}\times U(1)'_Y\right]_{SUSY}$$
$$\to SU(3)'_C\times SU(2)'_L\times SU(2)'_{\theta}\times U(1)'_X \times U(1)'_Z$$
\begin{equation} \to SU(4)'_C\times SU(2)'_L\times SU(2)'_{\theta}\times
U(1)'_Z\to SU(6)'\times SU(2)'_{\theta}\to E'_6 . \label{B} \end{equation} Now
we are confronted with the question: What group of symmetry
$SU(2)'$, unknown in the O-world, exists in the Sh-world, ensuring
the $E'_6$ unification?

\section{New shadow gauge group $SU(2)'$ and theta-particles}

In the present paper we consider the idea of the existence of 
theta-particles, developed by L.B. Okun \cite{9}. In those works 
it was suggested the hypothesis that in Nature there exists the 
symmetry group $SU(3)_C\times SU(2)_L\times SU(2)_{\theta}\times U(1)_Y$, 
i.e. with an additional non-Abelian $SU(2)_{\theta}$ group whose gauge 
fields are neutral, massless vector particles -- thetons, having a 
macroscopic confinement radius $1/\Lambda_{\theta}$.

We assume that the group of symmetry $G'_\theta = SU(3)'_C\times
SU(2)'_L\times SU(2)'_{\theta}\times U(1)'_Y$ exists in the
Shadow World at low energies instead of the SM'. By analogy with
theory \cite{9}, we have shadow thetons ${\Theta'}^i_{\mu\nu}$
($i=1,2,3$), which belong to the adjoint representation of
$SU(2)'_{\theta}$, three generations of shadow theta-quarks
$q'_{\theta}$, shadow leptons $l'_{\theta}$, and two theta-scalars
$\phi'_{\theta}$ as doublets of $SU(2)'_{\theta}$. Shadow thetons
have a confinement radius $1/\Lambda'_{\theta}$, and
$\Lambda'_{\theta}\sim 10^{-3}$ eV provides the tiny cosmological
constant. We also consider a complex scalar field
$\varphi_{\theta}$, which is a singlet under the symmetry group
$G'_\theta$. This singlet scalar field has its origin from 27-plet
of the $E'_6$ unification.

\section{The running of coupling constants in the O- and Sh-worlds}

In this work we consider the running of all the gauge
coupling constants in the SM and its extensions which is well
described by the one-loop approximation of the renormalization
group equations (RGEs), since from the Electroweak (EW) scale up
to the Planck scale ($M_{Pl}$) all the non-Abelian gauge theories
with rank $r\ge 2$ appearing in our model are chosen to be
asymptotically free. With this aim we consider only the Higgs
bosons belonging to the $N+\bar N$ representations for $SO(N)$ or
$SU(N)$ symmetry breaking (see \cite{7}).

The running of the inverse coupling constants are given by the
following expressions:
\begin{equation}
    \alpha_i^{(\prime)-1}(\mu) = \frac{b_i^{(\prime)}}{2\pi}\ln
    \frac{\mu}{\Lambda_i^{(\prime)}}, \label{1} \end{equation}
where $\mu$ is the energy scale. For compactness of notation, we
denote by $\alpha^{(\prime)-1}$ the inverse of various coupling
constants and by $X^{(\prime)}$ the various scales and values
belonging to either OW (the non-primed symbols) or ShW (the primed
symbols). In Eq.~(\ref{1}) $\alpha_i^{(\prime)} =
{(g_i^{(\prime)})}^2/{4\pi}$ and $g_i^{(\prime)}$ is the gauge
coupling constant of the gauge group $G_i^{(\prime)}$. Here
$i=1,2,3$ correspond to $U(1)$, $SU(2)$ and $SU(3)$ groups of the
$SM^{(\prime)}$. A big difference between the EW scales $v$ and
$v'$ will not cause the same difference between the gauge scales
$\Lambda_i$ and $\Lambda'_i$: $\Lambda'_i = \xi \Lambda_i$ with
$\xi \approx 1.5$ for $\zeta\approx 30$.

For the energy scale $\mu \ge M_{ren}^{(\prime)}$, where
$M_{{ren}}^{(\prime)}$ is the renormalization scale, we have the
following evolution for the inverse coupling constants given by
RGE in the one-loop approximation:
\begin{equation}
 \alpha_i^{(\prime)-1}(\mu) = \alpha_i^{(\prime)-1}\left(M_{ren}^{(\prime)}\right) +
 \frac{b_i^{(\prime)}}{2\pi}t^{(\prime)}, \label{2} \end{equation}
where $t^{(\prime)} =\ln\left(\mu/M_{ren}^{(\prime)}\right)$ is
the evolution parameter.

As an example of the evolutions (\ref{A}) and (\ref{B}) we have
used the following parameters: supersymmetric breaking scale in
the O-world $M_{SUSY}=10$ TeV, $\zeta =30$, i.e.
supersymmetric breaking scale in the Sh-world $M'_{SUSY}=300$
TeV, seesaw scales $M_R=M'_R=2.5\cdot 10^{14}$ GeV.

\begin{figure}[!b]
\includegraphics[height=58mm,keepaspectratio=true,angle=0]{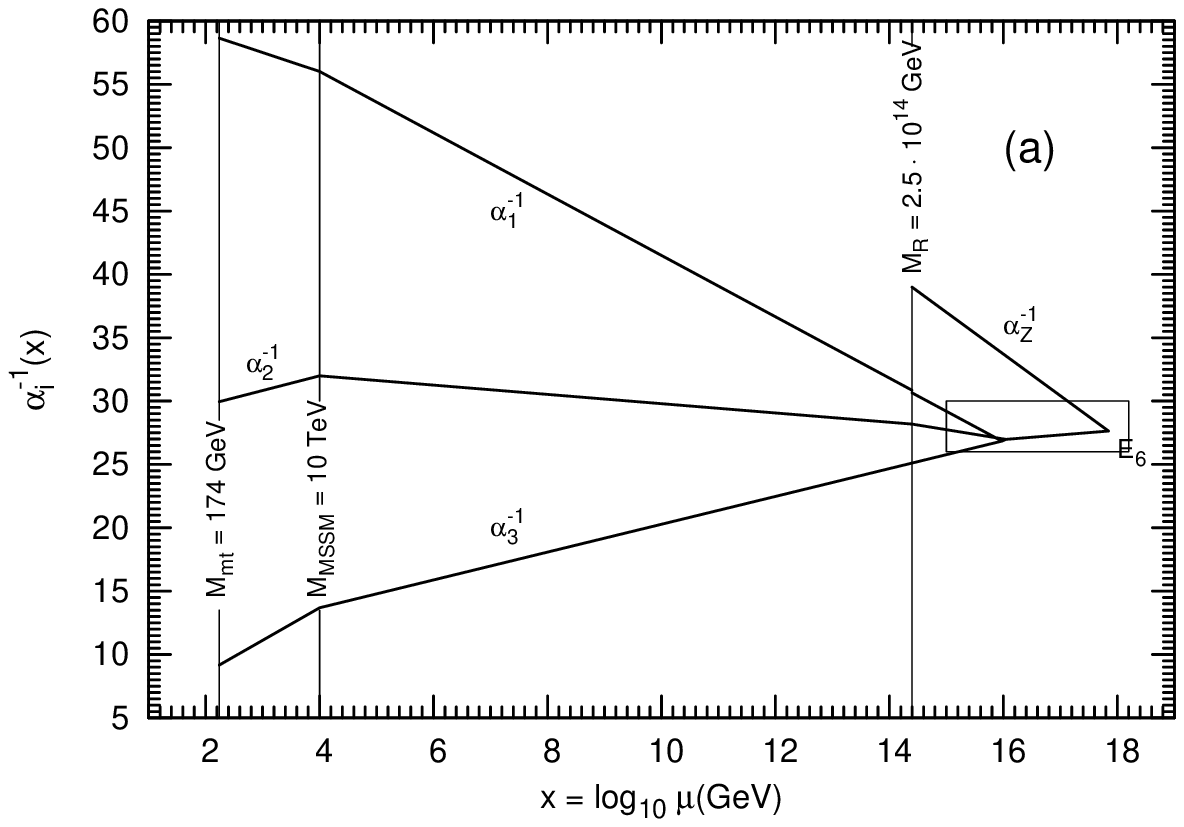}
\includegraphics[height=58mm,keepaspectratio=true,angle=0]{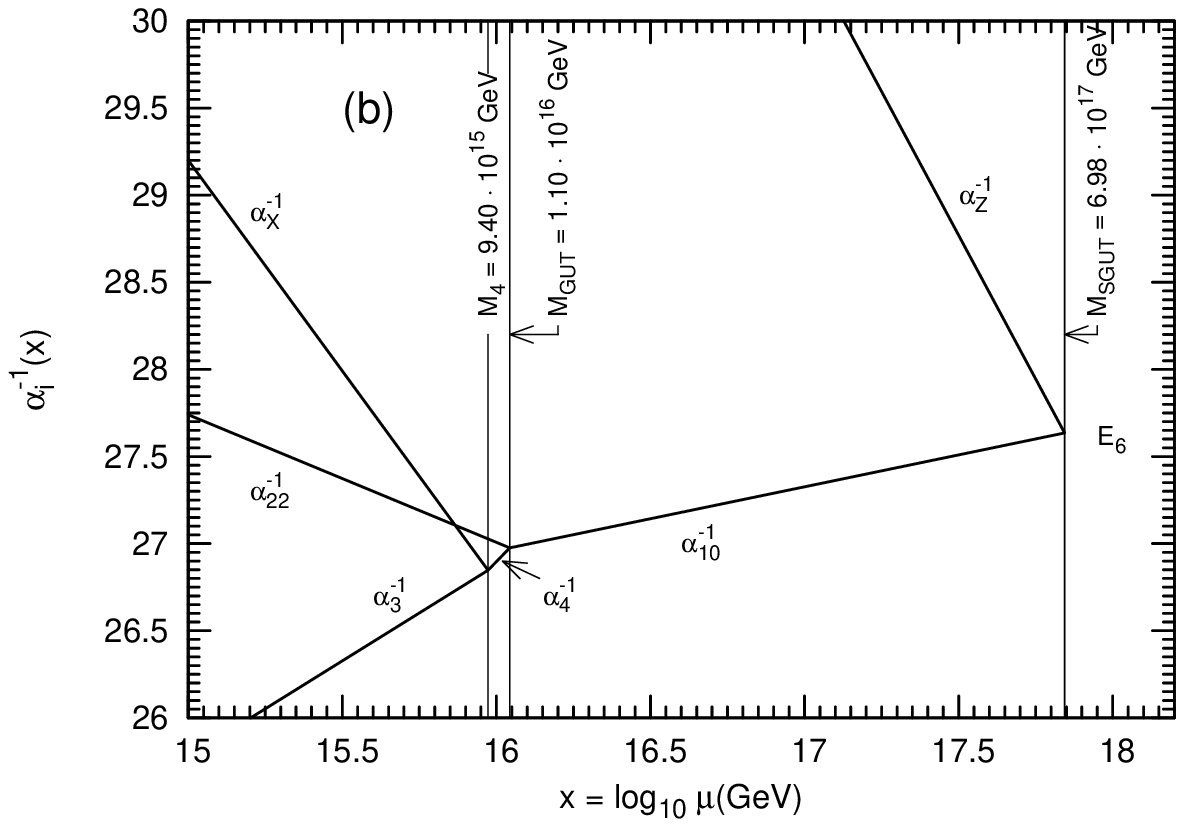}
\caption {Figure (a) presents the running of the inverse coupling
constants $\alpha_i^{-1}(x)$ in the ordinary world from the Standard
Model up to the $E_6$ unification for SUSY breaking scale $M_{SUSY}=
10$ TeV and seesaw scale $M_R=2.5\cdot 10^{14}$ GeV. This case
gives: $M_{SGUT}=M_{E_6}=6.98\cdot 10^{17}$ GeV and
$\alpha_{E_6}^{-1}=27.64$. Figure (b) is the same as (a),
but zoomed in the scale region from $10^{15}$ GeV up to the $E_6$
unification to show the details.}
\label{fig:1}
\end{figure}

\begin{figure}[!t]
\includegraphics[height=58mm,keepaspectratio=true,angle=0]{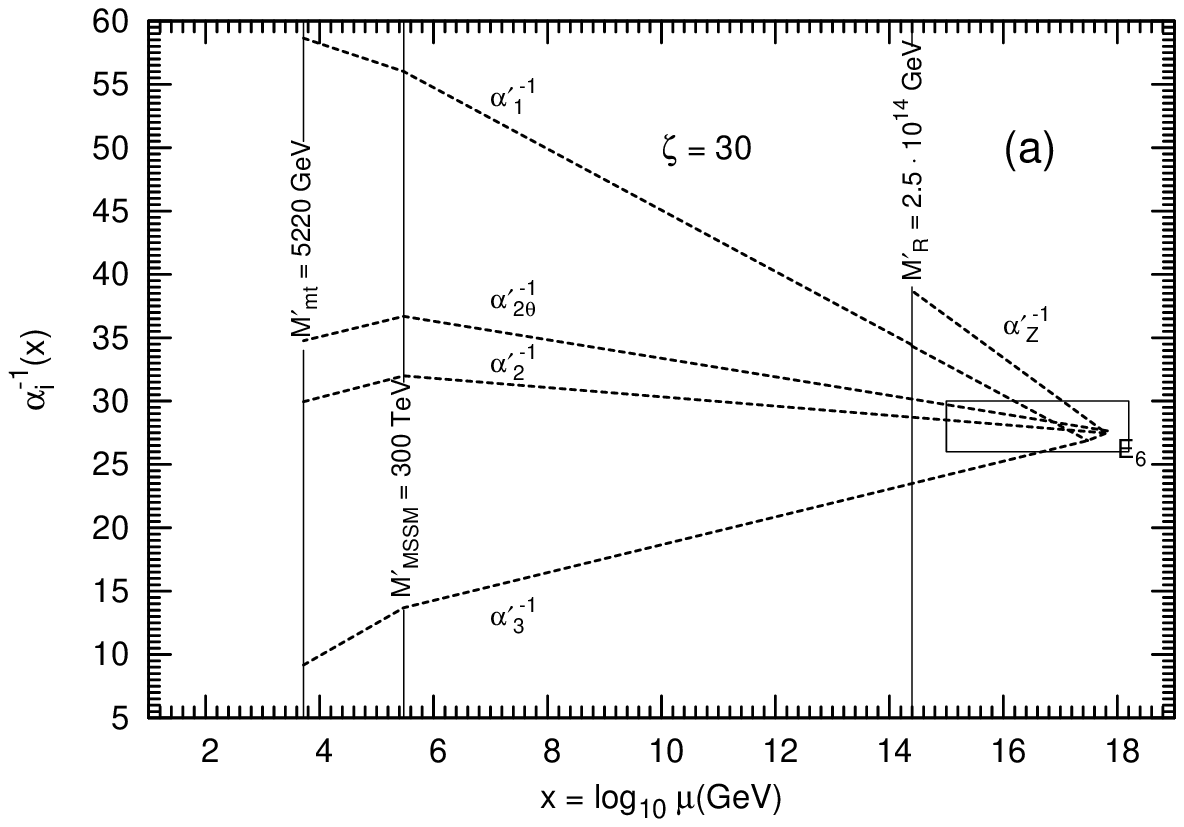}
\includegraphics[height=58mm,keepaspectratio=true,angle=0]{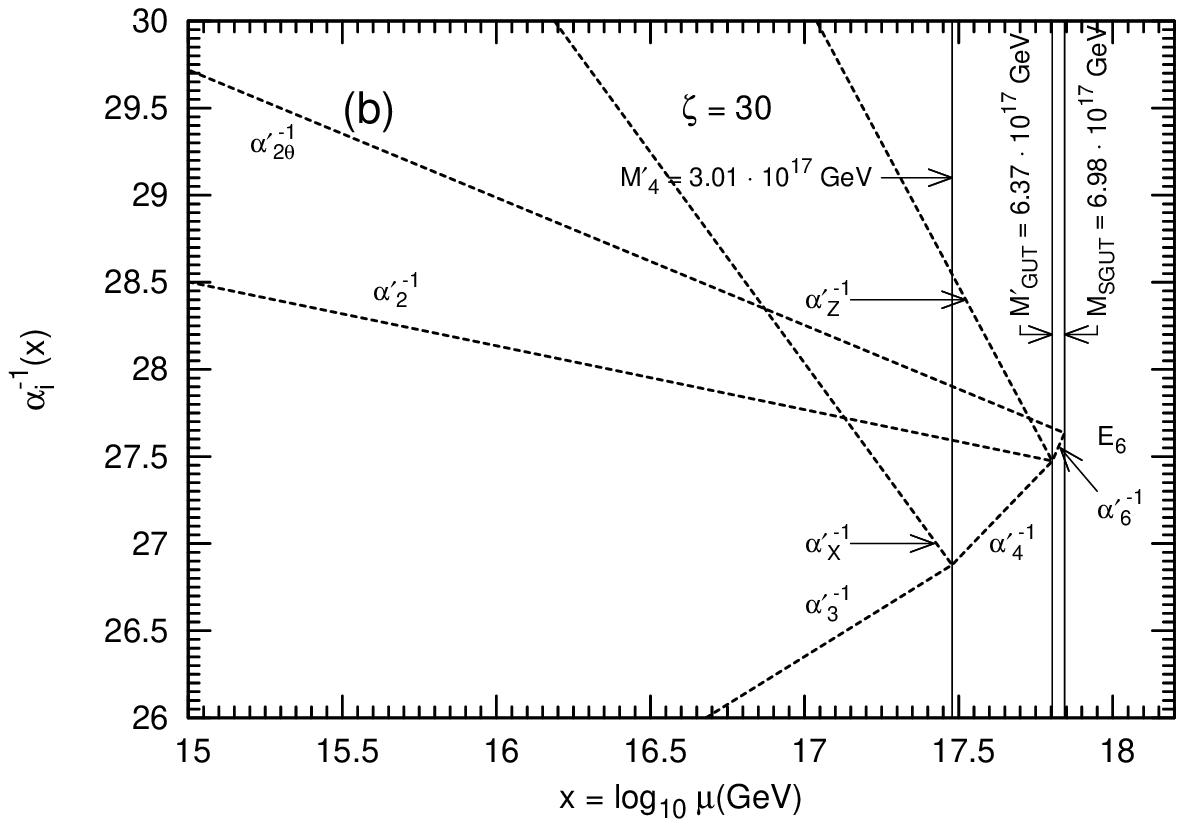}
\caption {Figure (a) presents the running of the inverse coupling
constants $\alpha_i^{-1}(x)$ in the shadow world from the Standard
Model up to the $E_6$ unification for shadow SUSY breaking scale
$M'_{SUSY}= 300$ TeV and shadow seesaw scale $M'_R=2.5\cdot
10^{14}$ GeV; $\zeta = 30$. This case gives: $M'_{SGUT}=M_{E_6}=6.98\cdot
10^{17}$ GeV and $\alpha_{E_6}^{-1}=27.64$. Figure (b) is the same
as (a), but zoomed in the scale region from $10^{15}$ GeV up to
the $E_6$ unification to show the details.}
\label{fig:2}
\end{figure}

The running of the inverse coupling constants as functions of
$x=\log_{10}\mu$ is presented for O-world in Fig. \ref{fig:1}(a,b) and for
Sh-world in Fig. \ref{fig:2}(a,b). We start in Fig. \ref{fig:1}(a) with $G_{SM}$ and
$M_{ren}=M_t$, where top-quark mass is $M_t=174$ GeV \cite{11}.
Fig. \ref{fig:2}(a) starts with $G'_{\theta}$ and $M'_{ren}=M'_t=\zeta
M_t=5.22$ TeV. In these pictures Figs. \ref{fig:1}(b),\ref{fig:2}(b) show the running of
the gauge coupling constants near the scale of the $E_6$
unification (for $x\ge 15$). The coefficients (slopes) $b_i$,
describing the running of the coupling constants with our choice
of gauge groups and particle content, are given in Tables \ref{tab:1},
\ref{tab:2}.

\begin{table}[!t]
\begin{tabular}{|r|c|c|c|c|}\hline
NonSUSY groups:&$SU(3)_C$&$SU(2)_L$   &$U(1)_Y$               &           \\
         $b_i$:&$7$      &$19/6$      &$-41/10$               &           \\\hline

   SUSY groups:&$SU(3)_C$&$SU(2)_{L,R}$&$SU(2)_L\times SU(2)_R$&$U(1)_Y$    \\
   $b_i^{SUSY}$:&$3$      &$-1$        &$b_{22}= -2$           &$-33/5$     \\

               &         &            &                       &           \\
               &$SU(4)$  &$U(1)_X$    &$U(1)_Z$               &$SO(10)$   \\
               &$b_4=5$  &$-33/5 $    &$-9$                   &$b_{10}=1$  \\\hline
\end{tabular}
\caption{The coefficients $b_i$ in the O-world with the breaking $E_6\to SO(10)\times U(1)_Z$.}
\label{tab:1}
\end{table}

\begin{table}[!t]
\begin{tabular}{|r|c|c|c|c|}\hline
NonSUSY groups: & $SU(3)'_C$ & $SU(2)'_L$ & $SU(2)'_{\theta }$ & $U(1)'_Y$ \\
                     $b_i'$: &     $7$      &    $19/6$    &    $3$    &  $-41/10$   \\\hline

  SUSY groups: & $SU(3)'_C$ & $SU(2)'_L$ & $SU(2)'_{\theta}$ & $U(1)'_Y$ \\
                     $b_i^{SUSY}$: &     $3$      &    $-1$      &     $-2$      &  $-33/5$    \\
                            &            &            &            &           \\
                            & $SU(4)'_C$  & $U(1)'_X$  & $U(1)'_Z$  & $SU'(6)$  \\
                            &     $5$      &    $-33/5$   &    $-9$      &   $11$      \\\hline
\end{tabular}
\caption{The coefficients $b_i$ in the Shadow World.}
\label{tab:2}
\end{table}

The running of ${\alpha'_{2\theta}}^{-1}(\mu)$ with slopes
$b_{2\theta} = 3$ and $b_{2\theta}^{SUSY}= -2$,
given by Fig. \ref{fig:2}(a,b), shows that it is easy to obtain the value
$\Lambda'_{\theta}\sim 10^{-3}$ eV.

\section{Shadow axion, cosmological constant and dark energy}

In the present paper we give a very simple explanation
for the smallness of the cosmological constant.

There exists an axial $U(1)_A$ global symmetry in our theory with
a current having $SU(2)'_\theta$ anomaly, which is spontaneously
broken at the scale $f_{\theta}$ by a singlet complex scalar field
$\varphi_{\theta}$, with a VEV $\langle \varphi \rangle =
f_\theta$, i.e.
\begin{equation} \varphi = \left(f_\theta + \sigma\right) \exp\left(i a_\theta/f_\theta\right).
\label{5} \end{equation}
The boson $a_{\theta}$ (imaginary part of the singlet scalar field
$\varphi_{\theta}$) is an axion and could be identified with a
massless Nambu-Goldstone (NG) boson if the $U(1)_A$ symmetry is
not spontaneously broken. However, the spontaneous breaking of the
global $U(1)_A$ by $SU(2)'_{\theta}$ instantons inverts
$a_{\theta}$ into a pseudo Nambu-Goldstone (PNG) boson.

A singlet complex scalar field $\varphi_{\theta}$ reproduces a
Peccei-Quinn (PQ) model \cite{10}. In the shadow world with shadow
$\theta$-particles the vacuum energy density is:
$\rho_{vac}={(\Lambda'_\theta)}^4$. Near the vacuum, a PNG mode
$a_\theta$ emerges the following PQ axion potential:
\begin{equation} V_{PQ}(a_\theta) \approx \left(\Lambda'_\theta\right)^4
\left(1 - \cos(a_\theta/f_\theta)\right).  \label{6} \end{equation}
This axion potential exhibits minima at \begin{equation}
\cos(a_{\theta}/f_\theta) = 1, \quad {\rm{or}} \quad
{(a_{\theta})}_{min}= a_n = 2\pi n f_\theta, \quad n = 0,1,...
\label{7} \end{equation} 
For small fields $a_\theta$ we expand the effective
potential near the minimum: 
\begin{equation}  V_{eff} \approx \left(\Lambda'_\theta\right)^4 \left(1 +
\frac 12 (a_\theta/f_\theta)^2 + ...\right) = {\left(\Lambda'_\theta\right)}^4
+ \frac 12 m^2 a_\theta^2 + ...,  \label{8} \end{equation}
and hence the PNG axion mass squared is given by: 
\begin{equation} m^2\sim {\Lambda'_{\theta}}^4/f^2_{\theta}.  
\label{9} \end{equation} 
Let us assume that at the cosmological epoch when $U(1)_A$ was 
spontaneously broken, the value of the axion field $a_\theta$ was 
deviated from zero, and it was $a_{\theta,in}\sim f_\theta$. The 
value of the scale $f_\theta \sim 10^{18}$ GeV (near the $E_6$ 
unification breaking scale) makes it natural that the $U(1)_A$ 
symmetry was broken before inflation, and the initial value 
$a_{\theta,in}$ was inflated above the present horizon. So after 
the inflation breaking scale, and in particular in the present 
universe, the field $a_\theta$ is spatially homogeneous (constant), 
and the initial energy density corresponding to $a_{\theta,in}$ is 
also spatially homogeneous:
\begin{equation} \rho_{in} = V(a_{\theta,in}) \simeq
\Lambda'^4_\theta \left(1 - \cos(a_{\theta,in}/f_\theta)\right).
\label{10} \end{equation} 
For the expanding universe the equation of motion
(EOM) of the classical field $a_\theta$ is: 
\begin{equation} \frac{d^2 a_\theta}{dt^2} + 3 H 
\frac{d a_\theta}{dt} + V'(a_\theta) = 0, \label{11} \end{equation} 
where $H$ is the Hubble parameter \cite{11}: $H = 1.5
\times 10^{-42}$ GeV. For small $a_\theta$ we have:
$V'(a_\theta)= m^2 a_\theta$. If $\Lambda'_\theta \sim 10^{-3}$ eV
and $f_\theta \sim 10^{18}$ GeV, then from Eq.~(\ref{9}) we obtain
a value of the axion mass: \begin{equation} m\sim
{\Lambda'_{\theta}}^2/f_{\theta}\sim 10^{-42} \,\,{\rm{GeV}}.
\label{13} \end{equation} 
Now, it is natural to assume that the initial velocity
$\dot{a}_{\theta,in}$ was small: $ \dot{a}_{\theta,in}\sim H
f_\theta$. Then, for $3H^2 \gg m^2$ the potential curvature
$V'(a_\theta)$ in the above EOM can be neglected, and we have a
solution with $a_\theta$ remaining the constant in time.

For the present epoch the critical density of the universe is: 
\begin{equation}\rho_{c} = 3H^2/8\pi G = {(2.5\times 10^{-12}\,\, 
{\rm{GeV}})}^4.\label{14} \end{equation} 
According to the Particle Data Group \cite{11}, the
fraction of the dark energy corresponds to 
\begin{equation} \rho_{DE} \approx
0.75 \rho_c \approx (2.3 \times 10^{-3} \,\, {\rm eV})^4. \label{15}
\end{equation} Now, having $m^2 < 3H^2$, we see that the classical PNG field
$a_\theta$ does not start the oscillation and in the present epoch
its energy density remains constant (does not scale with the time)
and saturates the dark energy fraction of the universe: 
\begin{equation}\rho_{DE}= \rho_{vac}= min\,\, V_{eff}\simeq \Lambda'^4_\theta,
\label{16} \end{equation} which means that 
$\Lambda'_\theta \approx 2.3 \times 10^{-3}$ eV.

In this case, for the present epoch, the energy of the PNG field
$a_\theta$ can imitate dark energy, providing the equation of the
state $\rho = w p$ with $w \approx -1$, but not exactly equal to
$-1$, as a quintessence. Of course, to claim that this can explain
the present amount of the dark energy, one must assume that the
major constant contributions to the cosmological term are
canceled by some means, i.e. true cosmological constant is almost
zero by some (yet unknown) symmetry, or by dynamical reasons. Also
the gravity itself can be modified so that it does not feel the
truly constant terms in the energy. In this case one can ascribe
the present acceleration of the universe by such a PNG
quintessence field, with implication that the acceleration will
not be forever, but it will finish as soon as $m^2 \sim 3H^2$ will
be achieved. After that the PQ classical energy will behave as a
dark matter component and not as a dark energy.

Here we have suggested a model when our universe was trapped in
the vacuum (\ref{16}), and exists there at the present time with a
tiny cosmological constant $CC$:
\begin{equation} CC=\rho_{vac}\simeq \left(\Lambda'_\theta\right)^4\simeq (2.3\times
10^{-3}\,\,\rm{eV})^4. \label{17} \end{equation}
Such properties of the present axion lead to the `$\Lambda CDM$'
model of the accelerating expansion of our universe \cite{11}.  By
this reason, the axion $a_{\theta}$ could be called an
`acceleron', and the field $\sigma$ given by Eq.~(\ref{5}) is an
`inflaton'.

\section{Dark matter} The existence of dark matter in the
universe, which is non-luminous and non-absorbing matter, is now
well established by astrophysics.

For the ratios of densities $\Omega_X = \rho_x/\rho_c$,
cosmological measurements give the following density ratios of the
total universe \cite{11}: $\Omega_0 = \Omega_r + \Omega_M +
\Omega_\Lambda = 1$. Here $ \Omega_r$ is a relativistic
(radiation) density ratio, and $\Omega_{\Lambda} = \Omega_{DE}$.
The measurements give: $\Omega_{DE}\sim 75\%$ - for the mysterious
dark energy, $\Omega_M \approx \Omega_B + \Omega_{DM} \sim 25\%$,
$\Omega_B \approx 4\%$ - for (visible) baryons, $\Omega_{DM}
\approx 21\%$ - for dark matter. Here we propose that a plausible
candidate for DM is a shadow world with its shadow quarks,
leptons, bosons and super-partners, and the shadow baryons are
dominant:
 $\Omega_{DM} \approx \Omega_{B'}.$
Then we see that
$\Omega_{B'} \approx 5\Omega_{B},$
what means that the shadow baryon density is larger than the
ordinary baryon density.

The new gauge group $SU(2)'_{\theta}$ gives the running of
${(\alpha')}_{2\theta}^{-1}(\mu)$. Near the scale
$\Lambda'_{\theta}\sim 10^{-3}$ eV, the coupling constant
$g'_{2\theta}$ grows infinitely. But at higher energies
Fig. \ref{fig:2}(a,b) show that this coupling constant is
comparable with the electromagnetic one. Here we would like to
emphasize that shadow quarks $q'_{\theta}$ of the first generation
are stable, and can participate in the formation of shadow
``hadrons'',  which can be considered as good candidates for the
Cold Dark Matter (CDM). So we have the two types of shadow
baryons: baryons $ b'$ constructed from shadow quarks $q'$ which
are singlets of $SU(2)'_{\theta}$, and baryons $b'_\theta$
constructed from the quark $q'$ and two shadow $\theta$-quarks
$q'_\theta$, in order to preserve $\theta$-charge conservation.
Then,
$\Omega_{B'} = \Omega_{b'+b'_\theta}\approx 5\Omega_{B}.$ We shall
study in detail the DM in a forthcoming communication.

The present work opens the possibility to specify a grand
unification group, such as $E_6$, from Cosmology.

\end{document}